\documentclass{aa}

\usepackage{graphicx}
\usepackage{txfonts}
\usepackage{newtxtext,newtxmath}
\usepackage{enumerate}
\usepackage{booktabs,tabularx}
\usepackage{threeparttable}
\usepackage{enumitem}
\usepackage{textcomp}
\usepackage{makecell}
\usepackage{bm}
\usepackage[hidelinks,colorlinks=true,linkcolor=blue,citecolor=blue]{hyperref}
\usepackage[dvipsnames]{xcolor}

\begin{document}

   \title{The impact of lossy data compression on the power spectrum of the high redshift 21-cm signal with LOFAR}
   \titlerunning{The impact of lossy data compression on the high redshift 21-cm signal power spectrum}
   \authorrunning{Chege et al.}

   \author{J. K. Chege
          \inst{1}
          \and
          L.V.E. Koopmans \inst{1}
          \and
          A. R. Offringa \inst{2}
          \and
          B. K. Gehlot \inst{1}
          \and
          S.A. Brackenhoff \inst{1}
          \and
          E. Ceccotti \inst{1}
          \and
          S. Ghosh \inst{1}
          \and
          C. Höfer \inst{1}
          \and
          F.G. Mertens \inst{1,3}
          \and
          M. Mevius \inst{2}
          \and
          S. Munshi \inst{1}
          }

   \institute{Kapteyn Astronomical Institute, University of Groningen, PO Box 800, 9700 AV Groningen, The Netherlands\\
              \email{chege@astro.rug.nl}
         \and
             Netherlands Institute for Radio Astronomy (ASTRON), Oude Hoogeveensedijk 4, 7991 PD Dwingeloo, The Netherlands \\
        \and
            LERMA, Observatoire de Paris, PSL Research University, CNRS, Sorbonne Université, F-75014 Paris, France
             }

   \date{Received XXX; accepted YYY}

 
  \abstract
   {Current radio interferometers output multi-petabyte-scale volumes of data per year making the storage, transfer, and processing of this data a sizeable challenge. This challenge is expected to grow with the next-generation telescopes such as the Square Kilometre Array which will produce a considerably larger data volume than current instruments. Lossy compression of interferometric data post-correlation can abate this challenge, but any drawbacks from the compression should be well understood in advance.} {Lossy data compression reduces the precision of data, introducing additional noise to the data. Since high-redshift (e.g., Cosmic Dawn or Epoch of Reionization) 21-cm studies impose strict precision requirements, the impact of this effect on the 21-cm signal power spectrum statistic is investigated in a bid to rule out unwanted systematics.}
   {Using observed visibilities datasets from the LOFAR telescope as well as simulated ones, we apply \textsc{dysco} visibility compression, a technique to normalize and quantize specifically designed for radio interferometric data. The power spectrum of these data is  analyzed, and we establish the level of the compression noise in the power spectrum in comparison to the thermal noise.  We also examine its coherency behavior by employing the cross-coherence metric. Finally, for optimal compression results, we compare the compression noise obtained from different compression settings to a nominal 21-cm signal power.}
   {From a single night of observation, we find that the noise introduced due to the compression is more than five orders of magnitude lower than the thermal noise level in the power spectrum. The noise does not affect calibration. Furthermore, the noise remains subdominant to the noise introduced by the non-linear calibration algorithm used as a result of random parameter initialization across different runs. The compression noise shows no correlation with the sky signal and has no measurable coherent component, therefore averaging down optimally with integration of more data. The level of compression error in the power spectrum ultimately depends on the compression settings.}
   {\textsc{dysco} visibility compression is found to be of insignificant concern for 21-cm power spectrum studies. Hence, data volumes can be safely reduced by factors of $\sim 4$ and with insignificant bias to the final power spectrum. Data from SKA-low will likely be compressible by the same factor as LOFAR, owing to the similarities of the two instruments. The same technique can be used to compress data from other telescopes, but a small adjustment of the compression parameters might be required.}
   

   \maketitle
%

\section{Introduction}
Over the past several decades, radio interferometers have been expanding in physical dimensions, transitioning into what is referred to as the large-N regime by incorporating an ever-increasing number of antennas. This trend is particularly pronounced in low-frequency instruments, where the relatively low cost of antenna components makes it economically viable to construct phased arrays consisting of hundreds of antennas. Such arrays include the Low-Frequency Array \citep[LOFAR;][]{Vanhaarlem2013}, the New Extension in Nançay Upgrading LOFAR \citep[NenuFAR;][]{Zarka2012}, the Murchison Widefield Array \citep[MWA;][]{Tingay2013} and the Hydrogen Epoch of Reionization Array \citep[HERA;][]{Deboer2017}. The forthcoming SKA-Low \citep{Braun2019, Mellema2013} will have more than $10^5$ antennas, a size that is considerably larger than that of current instruments.

An increased number of observing antennas that are correlated (or other correlated receiver systems such as tiles or stations), causes a substantially larger data volume. This is because the data output from an interferometer --- referred to as a visibility --- is a short-time-integrated cross-correlation of electric fields from each antenna pair \citep{Thompson2017}, therefore scaling as $\mathcal{O}(N^2)$, where $N$ is the number of correlated components. Moreover, to cater to different requirements, the visibilities are often recorded and written to disk at high temporal and spectral resolutions over long observation durations and large instantaneous bandwidths. Temporal resolution is preferable for the study of highly dynamic time-domain phenomena such as transient Radio-Frequency Interference (RFI; e.g. \citealt{Gehlot2024}) or ionospheric effects \citep[e.g.][]{Chege2022}. Similarly, observations with high spectral resolution are required, for example, in the study of radio spectral lines \citep[e.g.][]{Asgekar2013}. Some science cases such as deep all-sky radio surveys may require observations of large sky areas \citep[e.g.][]{Shimwell2017} while others require integration of large data volumes obtained from the same sky field over a long duration. An example of the latter is the observations targeting the cosmic 21-cm signal emitted during the Epoch of Reionization (EoR) and Cosmic Dawn \citep[e.g.][]{Paciga2013, Patil2017, Cheng2018, Bharat2019, Kolopanis2019, Li2019, Mertens2020, Trott2020, Hera2022, Hera2023, Munshi2024}. All these varying requirements result in considerable volumes of unprocessed observational data.

The output volumes from current interferometers have already grown into the petabytes scales\footnote{The LOFAR EoR Key Science Project project alone has $\approx5$ petabytes of archival data.} \citep{Sabater2015} and with the forthcoming SKA-low telescope, the data deluge is bound to upsurge into unprecedented exabyte volumes. This `big data' demands large storage spaces coupled with complex network architectures for prompt retrieval and transfer, sometimes over thousands of kilometers, before commencing any processing.  For this reason, data archival and management alone becomes a significantly expensive and sometimes limiting part of the science project. A viable solution capable of saving substantial storage resources and considerably mitigating the input/output (I/O) bottleneck could be the application of data compression techniques.

A compression method can be said to be either lossy or lossless. The former refers to a loss of information after data compression while the latter fully preserves the information. Lossless compression methods (e.g GZIP\footnote{\hyperlink{https://fits.gsfc.nasa.gov/https://www.gzip.org/}{https://fits.gsfc.nasa.gov/https://www.gzip.org/}}), while fully preserving every single data bit, rely on structured information which is scarce in noisy data. Hence, lossless methods achieve only few tens of percent compression factors on noisy data \citep{Lindstrom2017}.  On the other hand, lossy methods achieve much higher compression factors on noisy data, at the cost of losing some information. 

At high resolution, radio data is very noisy, drawing a preference for lossy compression methods as opposed to lossless ones. However, lossy compression, as with any other data transforming step, should preferably not bias the final science output in any significant way. This is especially the case for high-redshift 21-cm studies, as they target an extremely faint signal requiring the integration of a thousand hours for, say, the LOFAR telescope before a detection can be achieved \citep[e.g.][]{Mesinger2016}. 

Furthermore, this signal is buried under strong galactic and extragalactic foregrounds that are 3--5 orders of magnitude brighter \citep[e.g.][]{Jelic2008, Jelic2010}, exacerbating the challenge. High-redshift 21-cm studies thus aim for minimal systematic biases and errors \citep[e.g.][]{Barry2016}. Hence errors resulting from lossy visibility data compression should not have a detrimental impact on this already considerable challenge.

Several algorithms have already been developed for the specific purpose of compressing radio visibility data. These include BITSHUFFLE \citep{Masui2015}, implemented for lossless compression of integer data from the Canadian Hydrogen Intensity Mapping Experiment (CHIME) and reported to achieve compression of data to almost a factor of four. Several other compression methods have been developed for specific data formats such as the Flexible Image Transport System \footnote{\hyperlink{https://fits.gsfc.nasa.gov/}{https://fits.gsfc.nasa.gov/}} \citep[FITS,][]{Wells1981} file format and Astronomical Image Processing System\footnote{\hyperlink{http://www.aips.nrao.edu}{http://www.aips.nrao.edu}} \citep[AIPS, see e.g.][]{White2012}. For noisy complex visibilities data in the Measurement Set\footnote{\hyperlink{https://casa.nrao.edu/Memos/229.html}{https://casa.nrao.edu/Memos/229.html}} data format, the Dynamical Statistical Compression \citep[\textsc{dysco};][]{Offringa2016} tool was developed to perform lossy compression and shown to achieve a compression factor of four or more on LOFAR and MWA visibilities.

In this paper, we investigate the impact of lossy compression on visibilities in 21-cm observations data processing as a means of tackling excessive data volume challenges. While antenna-specific recorded voltages are normally quantized before correlation in well-understood procedures, compression of data after correlation and its effect in the case of the 21-cm signal observations remains largely unexplored. Previous work by \citet{Offringa2016} investigated the image-space effects when compressing visibility data with \textsc{dysco}. In this work, we examine the impact of data compression specifically on studies of the high-redshift 21-cm signals using the power spectrum method which is the conventional metric of 21-cm signal measurements in most current EoR studies. We quantify the compression noise added to the power spectrum and establish its behavior. Specifically, we tackle three pertinent questions: i) What is the level of compression noise compared to thermal noise?; ii) is compression noise incoherent?; iii) does compression noise affect calibration?

We describe Dysco, the compression tool used in this work, in Sect. \ref{sec:compression}, before describing the observation and simulation data used in Sect. \ref{sec:data}. The data processing methodology is then described in Sect. \ref{sec:processing}, before presenting the results in Sect. \ref{sec:results}, and our conclusions in Sect. \ref{sec:conclusions}.


\section{Dynamical Statistical Compression (Dysco)}
\label{sec:compression}
In this section, we briefly summarize the data compression tool used throughout this paper, namely \textsc{dysco}.
\textsc{dysco}\footnote{\hyperlink{https://github.com/aroffringa/dysco}{https://github.com/aroffringa/dysco}} was developed by \citet{Offringa2016} and it consists of a visibility compression algorithm and a \textsc{casacore}\footnote{\hyperlink{https://casacore.github.io/casacore}{https://casacore.github.io/casacore}} standard data storage manager that enables transparent storage of compressed data in the measurement set format. In this way, the compressed data can be written to disk and processing can proceed normally without any additional steps. \textsc{dysco} is already integrated into both LOFAR and MWA data pre-processing pipelines.

\textsc{dysco} compression is performed in two consecutive steps: a normalisation and a quantization step. The normalization ensures that the full dataset has a constant noise variance. The noise distribution in visibility data can vary across different antennas, polarizations, timesteps, and frequencies. Therefore, assumptions made during the normalization step regarding the noise distribution across the four dimensions will have an impact on the compression accuracy. For instance, the `row normalization' (R) method assigns a scaling factor per `row', which contains data from the same antenna and timestep, but different polarizations and frequencies. Due to the assumption of uniform variance across multiple polarizations and frequencies, the row-normalization method has been shown to perform significantly worse in the image space by adding much higher noise in comparison to the other available methods. Alternatively, \textsc{dysco} incorporates two more normalization methods, namely the `row-frequency' (RF) and the `antenna-frequency' (AF) normalization methods. The former is similar to the row normalization method but with an additional scaling factor per frequency channel. The method also stores the per-polarisation normalization factors separately. The latter uses a three-term normalization factor composed of a frequency channel factor and a factor for each of the two correlated antennas. The normalization here is also done independently of each timestep and polarisation.

After normalization, the data is encoded using a non-linear quantization scheme with dithering. The encoding is optimized for complex samples using a distribution with zero mean, in such a way that probable values are more accurately compressed than less probable values. The choice of such a distribution is also optimizable by the \textsc{dysco} user. The encoded values are finally converted to binary values and bit-packed using a chosen number of bits.

The compression bit-size, normalization, and quantization distribution parameters available for compression in \textsc{dysco} are listed in Table~\ref{tab:dysco_params}. The default values used by the Default Preprocessing Pipeline\footnote{\hyperlink{https://github.com/lofar-astron/DP3}{https://github.com/lofar-astron/DP3}} (\textsc{DP3}; \citealt{Diepen2018}, used extensively for LOFAR data analysis are AF normalization, a Gaussian distribution which is truncated at 2.5$\sigma$ (only the distribution that is used to compute the ideal encoding is truncated; actual visibilities are never truncated, as during the normalization it is made sure that all visibility values fit within the chosen distribution), and 10 bits. In this paper, we study whether these default settings are sufficient to compress LOFAR 21-cm signal data, or whether other settings are needed. We finally note that given uncompressed data is typically stored in 32 or 64-bit format, storing it in a 10-bit format leads to the earlier-mentioned substantially smaller data volumes. We note that the meta-data in the measurement sets are not compressed and some meta-data (i.e. scale factors for the RF and AF normalizations) are added, hence leading to a slightly lower compression factor compared to the simple ratio of bits per visibility.

\begin{table*}[!tbp]
    \centering
    \caption{Bit-size, normalization method, and quantization distribution parameters as used for \textsc{dysco} compression in \textsc{DP3}. While many other bit-sizes can be used, here we list the bit-sizes as recommended from prior tests. The default \textsc{DP3} values are displayed in bold.}
    \label{tab:dysco_params}
    \begin{tabular}{lccr} 
        \hline
        Bit-size & Normalization & Distribution & Expected compression factor \\
        \hline
        8 & Row-Frequency & 1.5$\sigma$ Truncated Gaussian & 6\\
        \textbf{10} & \textbf{Antenna-Frequency} & \textbf{2.5$\sigma$ Truncated Gaussian} & 4\\
        12 & Row & 3.5$\sigma$ Truncated Gaussian & 3.5\\
        16 & & Gaussian & 2.5\\
        & & Uniform\\
        & & Student's T\\
        \hline
        \hline
    \end{tabular}
\end{table*}

\begin{table}
\caption{Information on the two nights of real LOFAR HBA observations analysed in this work.} 
\label{tab:all_nights}
\centering   
\begin{threeparttable}
\begin{tabular}{l c c c}
\hline\hline 
Parameter & L246297 & L246309 \\ 
\hline    
   Observation Cycle & 2 & 2 \\ 
   UTC$^a$ start date-time & \makecell{2014-10-23 \\ 16:46:30} & \makecell{2014-10-16 \\ 17:01:41} \\
   LST$^b$ start-time [hour] & 19.3 & 19.1 \\
   Duration [hour] & 13.0 & 12.6 \\
   SEFD$^c$ estimate & 4294 & 4253 \\ 
   Number of stations$^d$ & 62 & 62 \\ 
   Frequency range (MHz) & 148-160 & 148-160 \\ 
   Frequency resolution$^e$ (kHz) & 12.2, 61.0 & 12.2, 61.0 \\ 
   Time resolution (s) & 2.0 & 2.0 \\ 
\hline       
\end{tabular}
\begin{tablenotes}
  \small
  \item[a] Coordinated Universal Time
  \item[b] Local Sidereal Time.
  \item[c] System Equivalent Flux density; as reported in \cite{Mertens2020}.
  \item[d] International stations not included.
  \item[e] Values corresponding to data with 15 and 3 channel per sub-band i.e. before and after frequency averaging.

\end{tablenotes}
\end{threeparttable}
\end{table}

\section{Data Acquisition}
\label{sec:data}

In this section, we describe the data used in the rest of the paper, which includes both real and simulated radio observations.

\subsection{Real observations}
The datasets used in examining the effects of lossy data compression were obtained with LOFAR \citep{Vanhaarlem2013}. LOFAR is a low-frequency radio interferometer and a pathfinder instrument for the Square Kilometre Array with a geographical footprint centered in the Netherlands and spreading out into multiple European countries. It can observe in two frequency bands using Low-Band Antennas (LBA; 10--90 MHz) and High-Band Antennas (HBA; 110--240 MHz), respectively.
The antennas are phased-up into stations, with the core consisting of 48 stations (24 stations each split into two separate stations) densely packed within a 2-km-wide area near the town of Exloo in Drenthe. An additional 14 stations are located further across the Netherlands while 14 others are located in different European countries. These are referred to as the remote and international stations, respectively. The core, remote and international stations, have maximum baselines of approximately 4, 120, and 2000 km, respectively.

In our analysis, we used a typical LOFAR HBA dataset from the Cycle 2 observing season, retrieved from the LOFAR Long Term Archive. In total, two nights of observation were used, spanning a duration of 12 hours per night and a 12 MHz bandwidth between 148--160 MHz. The raw data has a 12 kHz and 2 s frequency and time resolution, respectively. More information about this dataset is summarised in Table~\ref{tab:all_nights}.

\subsection{Simulations}
Besides using real LOFAR observation data, we complemented our study with simulated data. These simulations are based on the measurement set of the real observation L246297, listed in Table~\ref{tab:all_nights}, with the simulated data replacing the observed data for consistent data structure and properties. We simulated two sets of data which we refer to as simulation sets A and B, respectively. Table~\ref{tab:all_nights_sim} lists the dataset properties and compression settings used for the different simulations. 

Firstly, for simulation set A, we used all core and remote stations to simulate two observation datasets, both spanning the same 12-hour duration. The datasets had an identical foreground emission comprising compact extragalactic radio sources but a unique and independent noise realization 
We modeled an area of $10\time10$ degrees around the North Celestial Pole (NCP) using the brightest $\sim700$ sources. We also included the far-field Cygnus A and Cassiopeia A sources as they are bright enough to have a significant impact on the processing of the NCP field. More details on this simplified NCP model can be found in Brackenhoff et al. 2024 (in prep).

For simulation set B, we used the same foreground model as simulation set A, but varied the compression bit-size and normalization parameters. In contrast with simulation A where we used different noise realizations per dataset, in simulation B, we added the same noise realization to all datasets. Here, we also limited ourselves to including only the core LOFAR stations and a shorter observation duration of 6 hours. This reduced dataset is chosen for less memory usage and quicker computation.

All datasets generated from both simulations (A and B) included the instrumental beam attenuation effect and were carried out at the same time and frequency resolution as the real raw data before any averaging, as listed in Table~\ref{tab:all_nights}. The simulations were done using the \textsc{sagecal}\footnote{\hyperlink{https://github.com/nlesc-dirac/sagecal}{https://github.com/nlesc-dirac/sagecal}} algorithm \citep{Yatawatta2013, Yatawatta2015}. For each simulated dataset, its uncompressed version was used as the reference dataset.

\begin{figure}[!t]
\includegraphics[width=\linewidth]{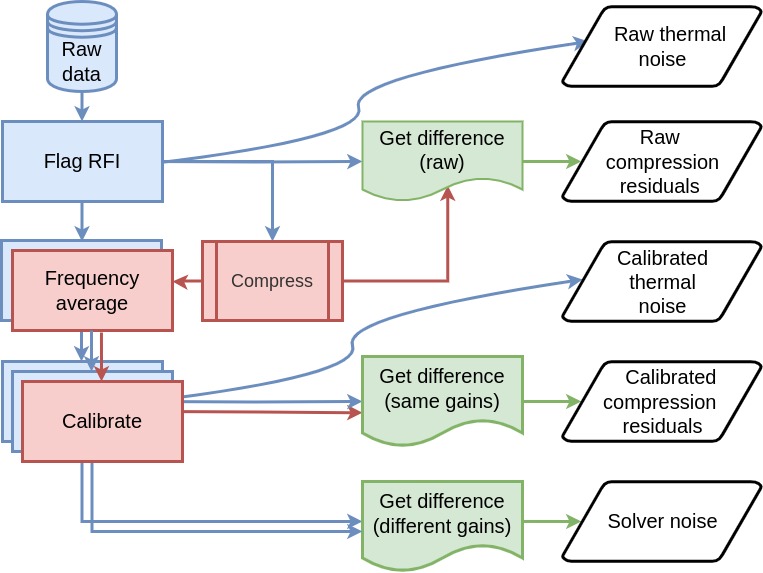}
\caption{Processing flow for the different visibility products used in obtaining the compression residuals noise, thermal noise and solver noise power spectra displayed in figures \ref{fig:1} and \ref{fig:1di}. The flow of compressed and uncompressed data is indicated in blue and red respectively, while different residuals are shown in green. Overlapping panels imply steps carried out multiple times on different or on the same dataset). The two curved lines represent obtaining thermal noise variance as the difference of even-odd timesteps image data. The exact details involved in each step are described in the text.}
\label{fig:workflow}
\end{figure}

\begin{table}[!ht]
\caption{Information on simulated LOFAR HBA observation data}
\label{tab:all_nights_sim}
\centering 
\begin{threeparttable}
\begin{tabular}{l c c c}
\hline\hline
Parameter & Simulation A & Simulation B \\ 
\hline
   MS template & L246297 & L246297 \\
   Stations & CS+RS & CS \\
   Duration [hour] & 12.0 & 6.0 \\
   Bit-size & 10 & 10, 12, 16 \\
   Normalization & AF & AF, RF \\
   Distribution & \makecell{$2.5\sigma$  truncated \\Gaussian} & \makecell{$2.5\sigma$  truncated \\Gaussian} \\
\hline                                   
\end{tabular}
\end{threeparttable}
\end{table}

\begin{figure*}
\includegraphics[width=\linewidth]{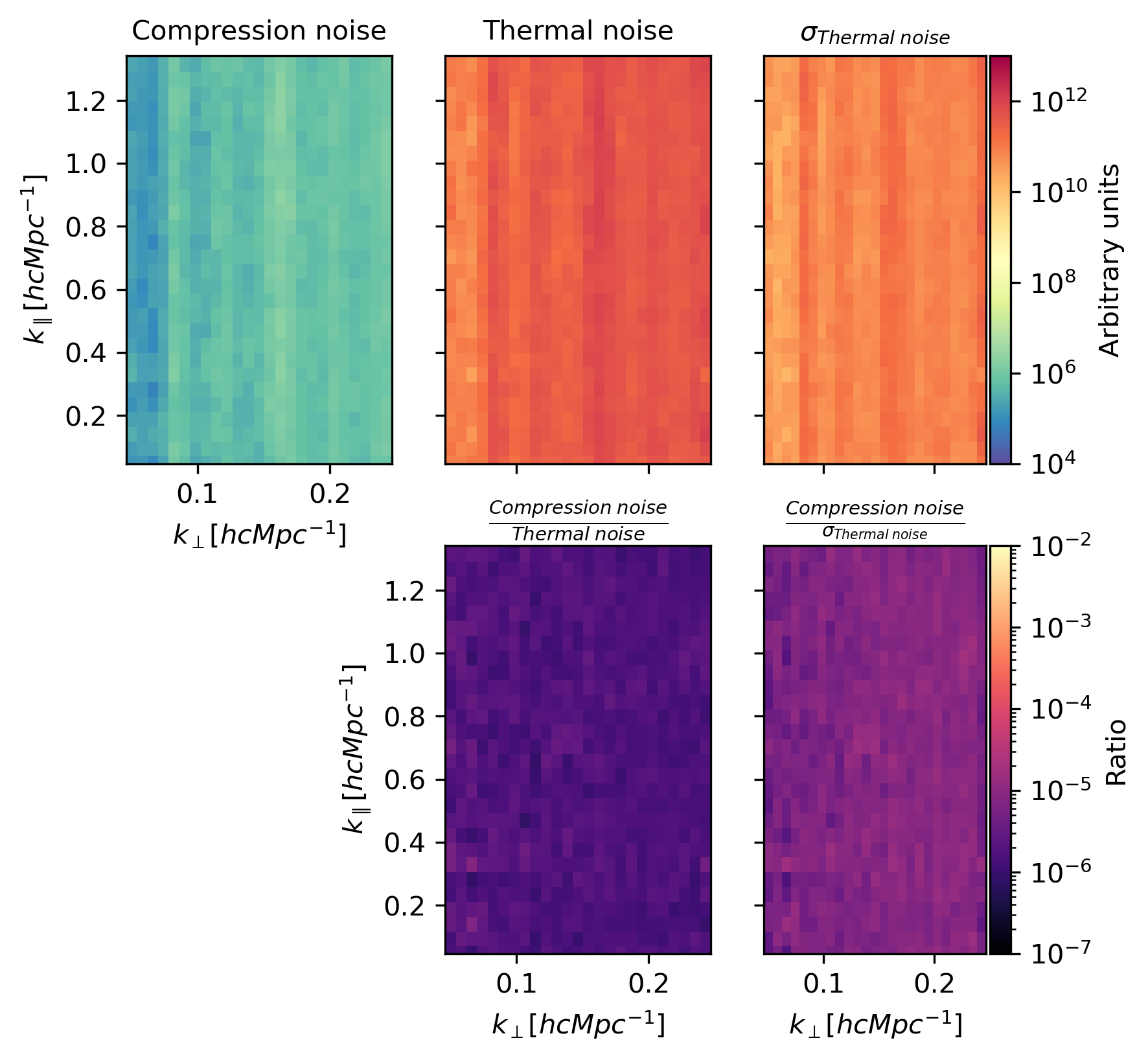}
\caption{Cylindrically-averaged power spectra comparing the compression noise (the compressed minus uncompressed visibilities) power spectrum to the thermal noise power spectrum. In the top row, the compression noise power spectrum (left), the thermal noise (middle) and the thermal noise uncertainty (right) are shown. In the bottom row, we show the ratio between the compression noise and both the thermal noise (middle) as well as the thermal noise uncertainty (right). These spectra are obtained from real uncalibrated data hence the arbitrary power spectrum units}
\label{fig:1}
\end{figure*}

\begin{figure*}
\includegraphics[width=\linewidth]{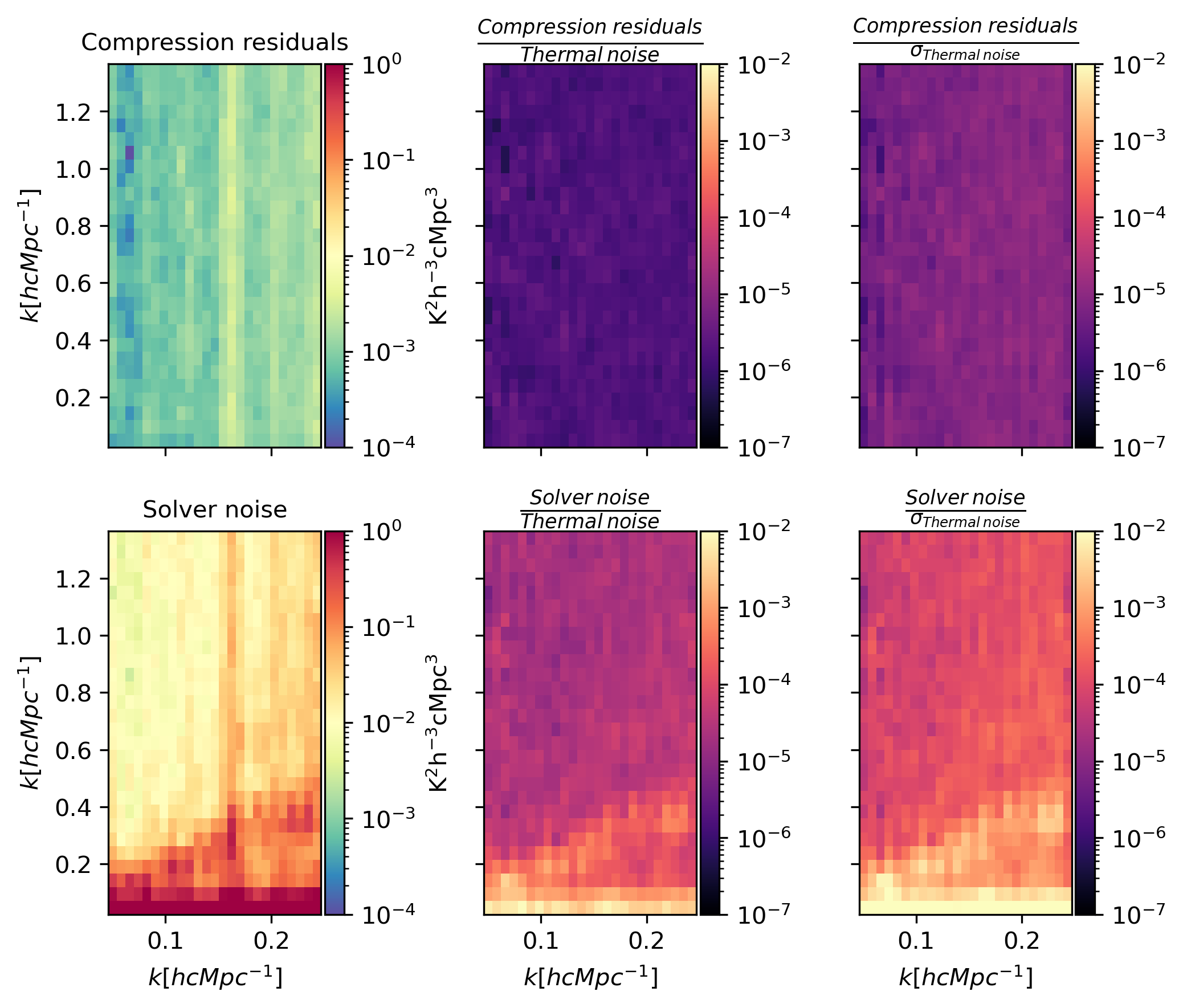}
\caption{Comparisons between the cylindrically-averaged power spectra of both compression and solver noise to the thermal noise and the thermal noise uncertainty after calibration. The top row shows the cylindrically-averaged power spectrum of the compression noise (top left) and its ratio with the thermal noise power spectrum (top middle) and the thermal noise uncertainty (top right). Similarly, the bottom row shows the solver noise power spectrum (bottom left) and its ratio with both the thermal noise power spectrum (bottom middle) and the thermal noise uncertainty (bottom right). The thermal noise and compression residuals here are obtained from real calibrated data.}
\label{fig:1di}
\end{figure*}

\section{Data Processing, Compression and Calibration}
\label{sec:processing}
In this section, we describe the data reduction steps carried out for our analysis. The analysis closely follows the steps applied in the LOFAR EoR Key Science Project data processing pipeline \citep[see][]{Mertens2020}. However, we introduce data compression as an additional data pre-processing step where needed.

\subsection{Pre-processing and compression}

In the pre-processing step, data was first run through an RFI excision step carried out using AOFlagger\footnote{\hyperlink{https://sourceforge.net/projects/aoflagger}{https://sourceforge.net/projects/aoflagger}} \citep{Offringa2012}. This step serves not only to get rid of unwanted terrestrial signals but also to reduce any dynamic range contribution by RFI to the data. A reduced dynamic range improves the performance of the normalization step during compression improving the overall compression performance. While RFI flagging has no effect on simulated data, we retained the RFI flags obtained from real data in our simulation datasets to replicate a realistic observation. All data from the international stations was also flagged in this step.

The data was then compressed using \textsc{Dysco}. We first used the default \textsc{DP3} compression parameters recommended by \cite{Offringa2016} for our relatively high-resolution noise-dominated data. Such noisy data is typical for 21-cm signal studies since they usually target sky fields with minimal foreground power and ease of foregrounds modeling. Later, we examine whether these default parameters are sufficient for the precision required in high-redshift 21-cm signal studies. For improved signal-to-noise ($\text{S/N}$-ratio) ratio per solution time interval during calibration and for quicker computation, the data was averaged to 61 kHz per spectral channel.

\subsection{Calibration}

Throughout this work, we limited ourselves to performing the first direction-independent (DI) calibration step in the LOFAR EoR pipeline.
One calibration step is sufficient for answering the question of whether compression errors affect calibration and its results would apply to the other stages of calibration. This is because compression is only applied once on the highest resolution raw data as it is the most voluminous. This is typically done for archival purposes. Decompression is then done prior to any further processing and therefore any compression effects on calibration should manifest clearly in the first calibration stage. Moreover, multiple compression and decompression runs during processing are not recommended as each iteration would introduce additional noise to the visibilities. Averaging in time and frequency carried out during processing reduces the data volume by a factor of 25 (the data is averaged from 15 to 3 frequency channels per sub-band for direction independent and from 2 s to 10 s time integration for the direction-dependent calibration step). However, intermediate visibilities during the different calibration steps occupy additional columns in the measurement set albeit requiring relatively limited disk space due to the averaging.

All gain calibration was carried out using \textsc{sagecal} \citep{Yatawatta2016}. For the real data, the calibration sky model was composed of two sky directions, one around the NCP and the other for the bright source 3C61.1 with 1333 and 1545 components, respectively. A calibration solution interval of 30 s was used with a single solution being obtained for each 183.3 kHz sub-band. A minimum and maximum calibration baseline cutoff was set at $50\lambda$ and $5000\lambda$, respectively, and the resulting gains were regularized using a third-order Bernstein polynomial. The resulting Bernstein polynomial for the central NCP direction was applied to the data to obtain the calibrated visibilities. The baseline cutoff and gains smoothing have been deemed crucial to minimize signal suppression and noise boost in the 21-cm power spectrum \citep[e.g.][]{Barry2016, Mevius2022}. For more details on these calibration parameter choices, see \citet{Patil2017, Sardarabadi2019} and \citet{Mertens2020}. The simulated data was calibrated similarly, with the only difference being in the sky model which in this case was composed of fewer components since the simulations had fewer compact foregrounds.

\subsection{Generation of power spectra}
Results presented in this paper are based on power spectra generated from different visibility datasets, real or simulated, raw or calibrated using the \textsc{pspipe}\footnote{\hyperlink{https://gitlab.com/flomertens/pspipe}{https://gitlab.com/flomertens/pspipe}} tool. In this section, we describe the process involved.

First, the visibilities from all sub-bands are gridded and transformed into image cubes using WSClean \citep{Offringa2014}. Each image cube has a field-of-view (FOV) of $12^{\circ}\times12^\circ$ centered at the NCP with an angular resolution of $0.5$ arcmin. For the actual power spectrum, this FOV is then reduced to $4^{\circ}$ by use of a Tukey spatial window. The image cube is then regridded and converted from image space units of Jy/PSF\footnote{PSF here abbreviates the Point-Spread Function.}, into brightness temperature units of Kelvins together with a spatial Fourier transform step back into visibilities space. This final step is carried out on a visibilities subset including only baselines between $50\lambda$ and $250\lambda$ in length. Concurrently, an estimate of the thermal noise variance is obtained by generating a new cube composed of the difference between even and odd timesteps. From such cubes, we can obtain the power spectrum by first taking a Fourier transform along the frequency direction. The coordinates are then mapped into comoving distances in the form of wavenumbers ($\mathbf{k}$) with the appropriate cosmological units \citep{Morales2004, McQuinn2006}. We use the common cylindrically-averaged (2D; $k_{\perp}$ and $k_{\parallel}$ coordinates) and the dimensionless spherically-averaged (1D, $k$) power spectra.

\begin{figure*}
  \centering
  \includegraphics[width=\linewidth]{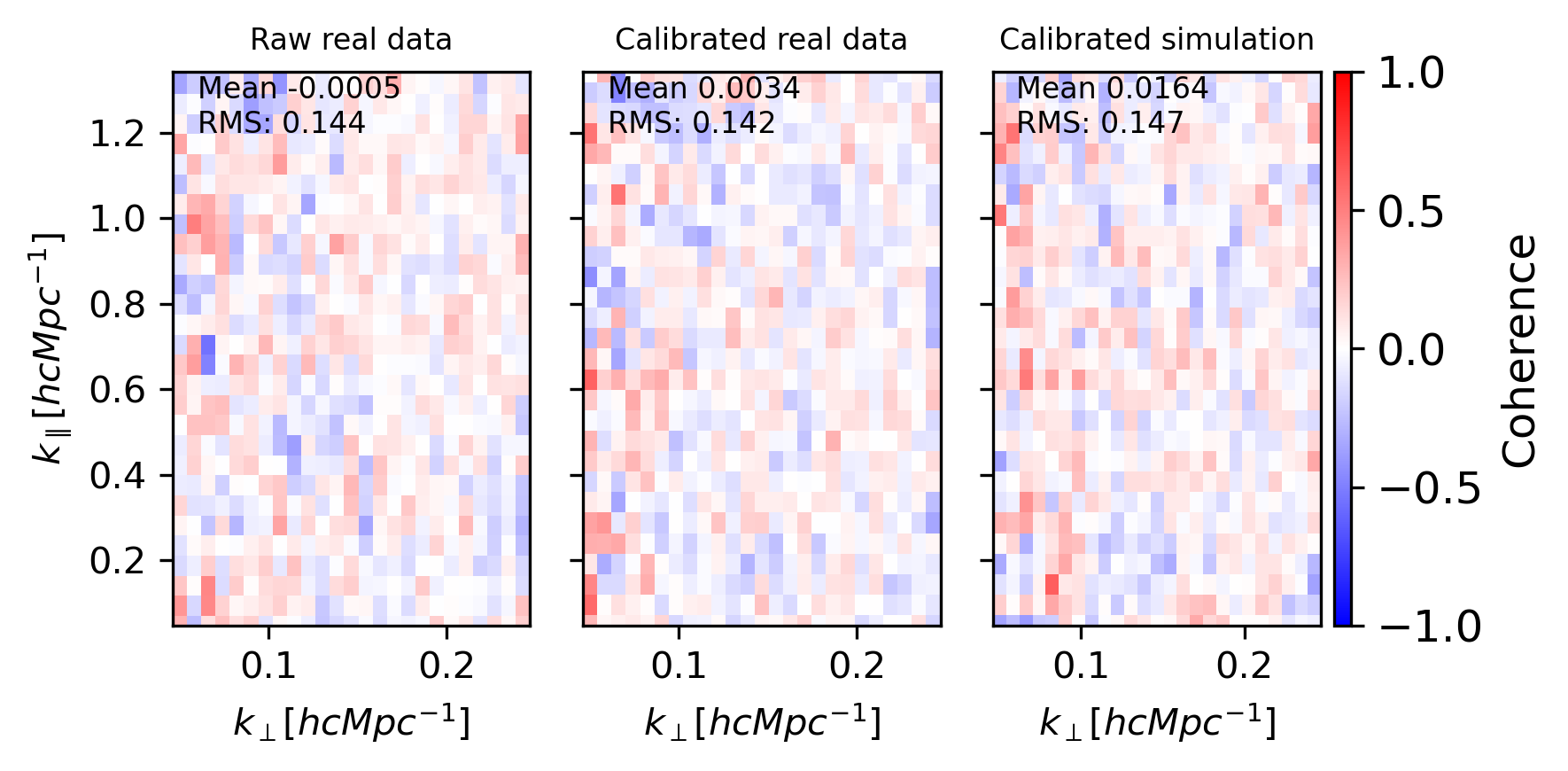} \\
  \includegraphics[width=\linewidth]{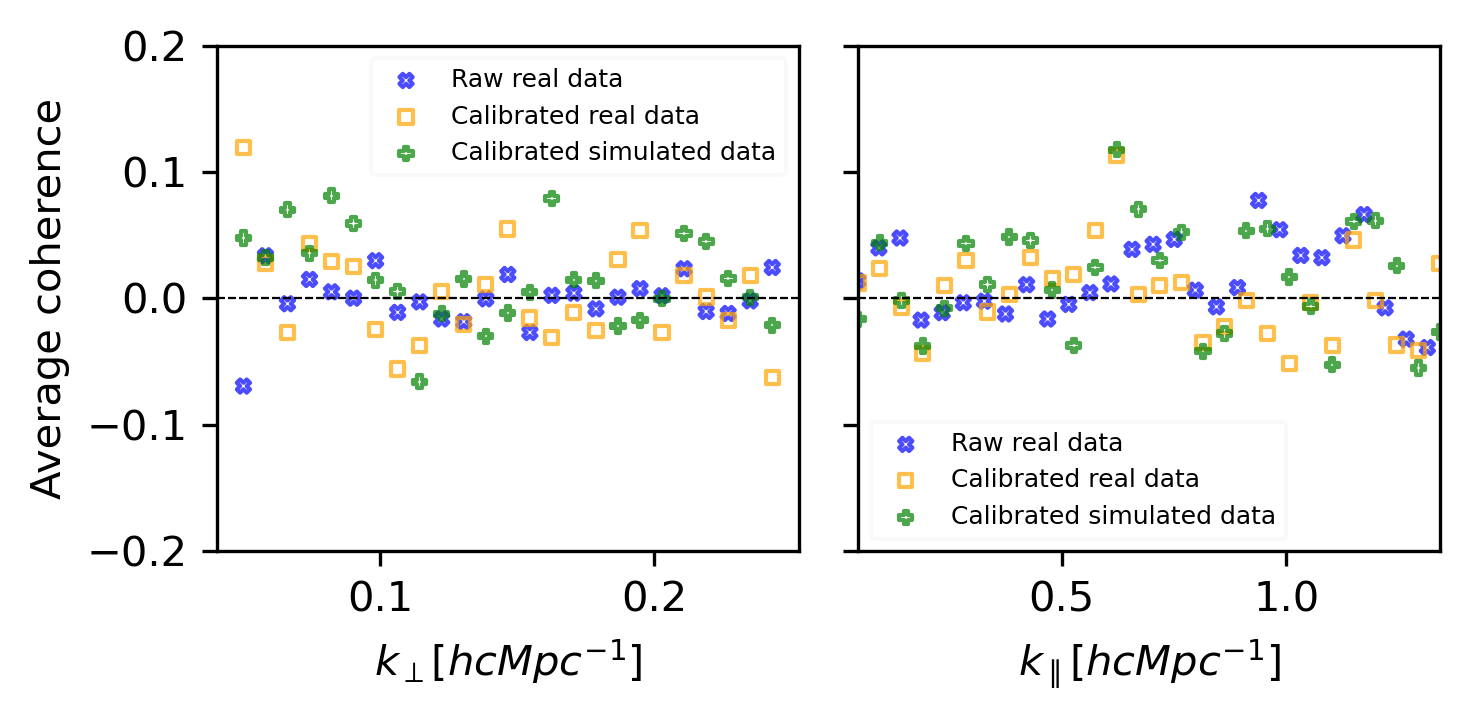}
  \caption{Compression residuals coherence for raw real, calibrated real and calibrated simulated data.  For each panel in the top row, the coherence is computed using two compressed minus uncompressed data residuals from two separate datasets. To eliminate solver noise, the calibration gains solution from each uncompressed dataset are applied to its compressed version before obtaining the residuals. The average coherence for each top row panel is shown in bottom row as a function of both $k_\perp$ (left) and $k_\parallel$ (right) modes.} 
    \label{fig:coherence_2dps_and_average}
\end{figure*}

\section{Results}
\label{sec:results}

In this section, we present the results of our analysis. First, we establish the scale of the compression noise in comparison to thermal noise. We then show the coherence properties of the compression noise followed by an assessment of how data compression affects calibration. Finally, we examine what are the optimal \textsc{dysco} compression settings.

\subsection{Compression noise}
\label{subsec:compression_noise}
To determine the additional noise introduced to the 21 cm power spectrum due to visibility compression, we computed the difference between the compressed and the reference (not compressed) visibilities before calibration.
Throughout the following sections, we will refer to the output of this subtraction as the compression residuals or the compression noise (Fig. \ref{fig:workflow}). Figure \ref{fig:1} shows the comparison between these compression residuals and the reference thermal noise by use of the cylindrically-averaged power spectrum. The noise introduced to the power spectrum due to data compression is shown to be around $5.5$ orders of magnitude lower than the raw thermal noise. Similarly, the compression noise power is shown to be $4.5$ orders of magnitude lower than the uncertainty of the reference thermal noise. Hence, even if compression noise were fully coherent (e.g., the result of compression of coherent foreground emission), it would only reach the level of the error on the thermal noise by adding about $30000$ times more data. Such a huge amount of data will, most likely, never be a requirement. Additionally, the ratios are devoid of any spatial structures, implying that compression noise does not introduce any spurious or scale-dependent errors. We will study the coherence shortly, showing that the compression noise is consistent with being incoherent.

A similar metric can be obtained for calibrated data by applying identical calibration gains solutions to both the reference and the compressed dataset before obtaining the calibrated compression residuals. The need for identical solutions is to eliminate calibration `solver noise', a term which refers to the additional noise introduced due to random initialization of parameters by \textsc{sagecal} per calibration run, which leads to slightly different gain solutions after a finite number of iterations during the optimization \citep[e.g.,][]{Mevius2022}. It is known that calibration also introduces a systematic power contribution resulting from various factors, for example, the use of incomplete sky models, transfer of gains solution from longer to shorter baseline sets, and spectrally-noisy gain solutions(e.g. \citep[e.g.][]{Barry2016, Mevius2022}). This \textit{systematic} power is different from the solver noise referred to here, which is indeed \textit{random} and intrinsic to the calibration algorithm used. Without identical calibration gains, the compression residuals would be dominated by this solver noise, although the latter is well below the thermal noise as shown below. We thus apply the DI calibration gains obtained for the reference dataset to its compressed version (it will be shown in Sect. \ref{sec:calibration_effects} that the results do not change significantly if calibration solutions obtained from the compressed dataset are applied instead).

Subsequently, we reran the power spectrum computation on the calibrated compression residuals, and the result is shown in Fig. \ref{fig:1di}. Similar to raw data, the compression of visibilities does not introduce spurious noise after calibration. The ratio between the compression residuals and the thermal noise level remains at orders lower than $10^{-5}$ between their 2D power spectra. A slightly higher level is seen in the ratio of the compression residuals and the thermal noise uncertainty. Additionally, solver noise bias is shown to be more dominant in comparison to the compression residuals noise, but also well below the thermal noise and the error on the thermal noise.

\begin{figure}
    \includegraphics[width=\linewidth]{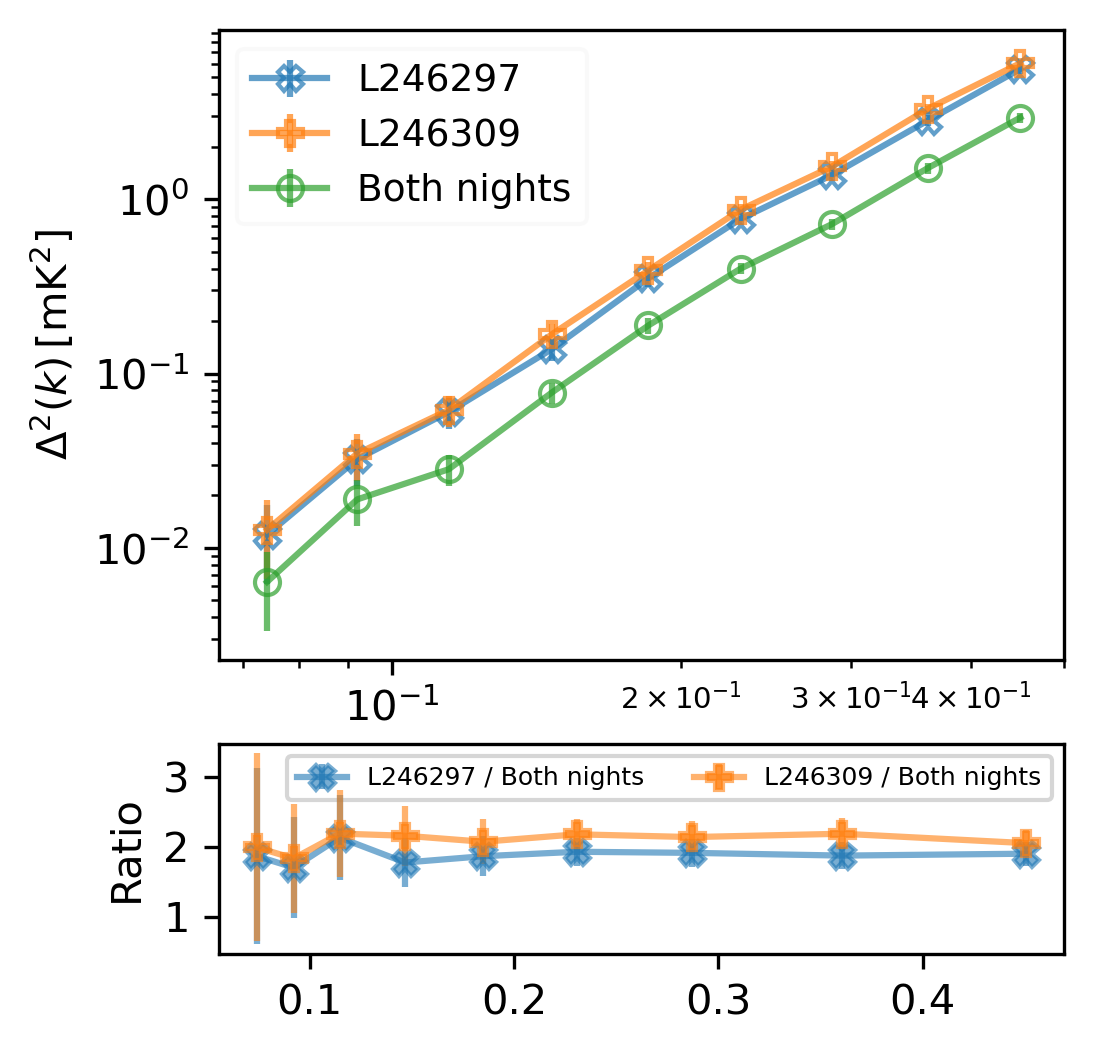}
    \caption{Spherically-averaged power spectra obtained from the compression residuals of two nights and their combined power spectrum. The ratio of each night to the combined power spectrum is shown in the bottom plot.}
    \label{fig:combined_compression_noise_3dPS}
\end{figure}

\subsection{Compression noise coherence}
\label{subsec:coherence}
Having determined that the added compression noise is far below both the thermal and solver noise for LOFAR-HBA data, we investigate whether this noise has any correlation with the sky signal and therefore is coherent in nature. Partly coherent compression noise would not only average down slower than incoherent noise, but it could also introduce biases on the 21-cm measurements obtained from deep integrations. For this test, we obtained three pairs of compression residuals from:

\begin{enumerate}[align=left]
    \item Both LOFAR nights before calibration,
    \item Both LOFAR nights after calibration, and
    \item Two identical simulations with different noise realizations (simulation A).
\end{enumerate}

As summarized in Table~\ref{tab:all_nights_sim}, all the datasets used for simulation A spanned an equal 12-hour duration, with the same LST range. The simulated LOFAR HBA datasets comprised of identical extragalactic foregrounds but a different thermal noise realization was added to each.
We processed all the datasets similarly, again applying the gains obtained from each uncompressed dataset to its compressed version, in order to get rid of the solver noise. We then obtain the difference of the DI-calibrated visibilities for each pair of reference and compressed simulated data resulting in a pair of DI-calibrated compression residuals. It is on each of this trio of residuals pairs that we check for any correlation.

We use the coherence metric $C$, given by the real part of the normalized cross-power spectrum \citep{Mertens2020, Gehlot2024}.

\begin{equation}
    \label{eqn:coherence_real_part_cross_ps}
    C_{a, b} \left(k_{\perp}, k_{\parallel} \right) = \frac{ \Re{ \left( \Tilde{T}_{a}^{*}(\mathbf{k}) \Tilde{T}_b(\mathbf{k}) \right )} }{ \sqrt{ |\Tilde{T}_a(\mathbf{k})|^2 |\Tilde{T}_b(\mathbf{k})|^2 } }
\end{equation}

Taking the real values of the cross-spectrum as opposed to the absolute values used in \cite{Mertens2020}, provides information on how positive and negative coherence values are distributed around zero. Therefore, this metric ranges from $-1$ to $1$ with both extremes denoting maximum coherence while zero denotes total incoherence. Although the coherence can have an imaginary component, due to spatial shifts between modes before and after compression, such an effect would require compression to be highly spatially correlated which is not the case. We can therefore ignore the imaginary component.
The coherences from each pair are shown in the top panels of Fig. \ref{fig:coherence_2dps_and_average}. The compression noise is seen to be highly incoherent across all $k$-modes with a noise-like behavior around a mean of zero, devoid of any spurious coherence structures. The coherence has an rms of $\sim{0.14}$ around a mean of zero which remains consistent across all three cases. In the bottom panel, we present the average coherence with respect to both the $k_\perp$ and $k_\parallel$ modes. The average coherence again has a noise-like behavior around zero that is within the rms and consistent for both cases.

To ascertain that this coherence level is consistent with random uncorrelated residuals data, we computed the spherically-averaged power spectrum of the compression residuals from each night separately and then compared it with the power spectrum obtained from a coherent averaging of the compression residuals from both nights. Figure \ref{fig:combined_compression_noise_3dPS} shows the three spectra as well as a ratio of each individual night's residual power spectrum to the combined power spectrum. Both ratios show a consistent factor of $\sim2$ as expected from combining two equal-size datasets composed of highly incoherent noise. This verifies that compression noise will progressively average down, like normal system noise, with deeper data integrations. Any coherence in the compression noise would not rise above the error on the thermal noise  for at least several hundred thousand hours of integration.

For completeness, we also examine the correlation of the solver noise as obtained from the calibration of two different observation nights. This noise is also found to have minimal coherence with a mean of --$0.013$ and an RMS of 0.18 as shown in Fig. \ref{fig:solver_noise_coherence}. This too, while not being the main subject of this paper, is a novel result. It shows that the random algorithmic solver noise, examined in this work using \textsc{sagecal}, does not introduce significant bias in the power spectrum.

\subsection{Calibration on compressed data}
\label{sec:calibration_effects}
\begin{figure}
    \includegraphics[width=\linewidth]{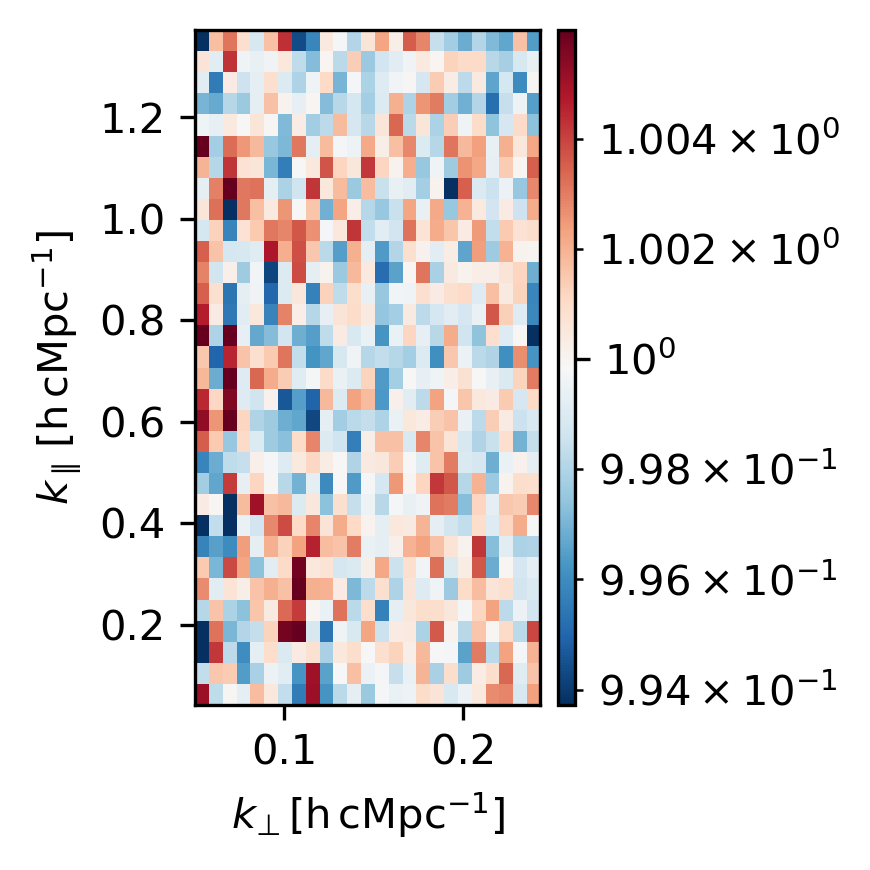}
    \caption{The ratio of compression residuals after applying either the gains solutions obtained from uncompressed data or the ones obtained from compressed data.}
    \label{fig:simulations_residuals_coherence}
\end{figure}

\begin{figure}
    \includegraphics[width=\linewidth]{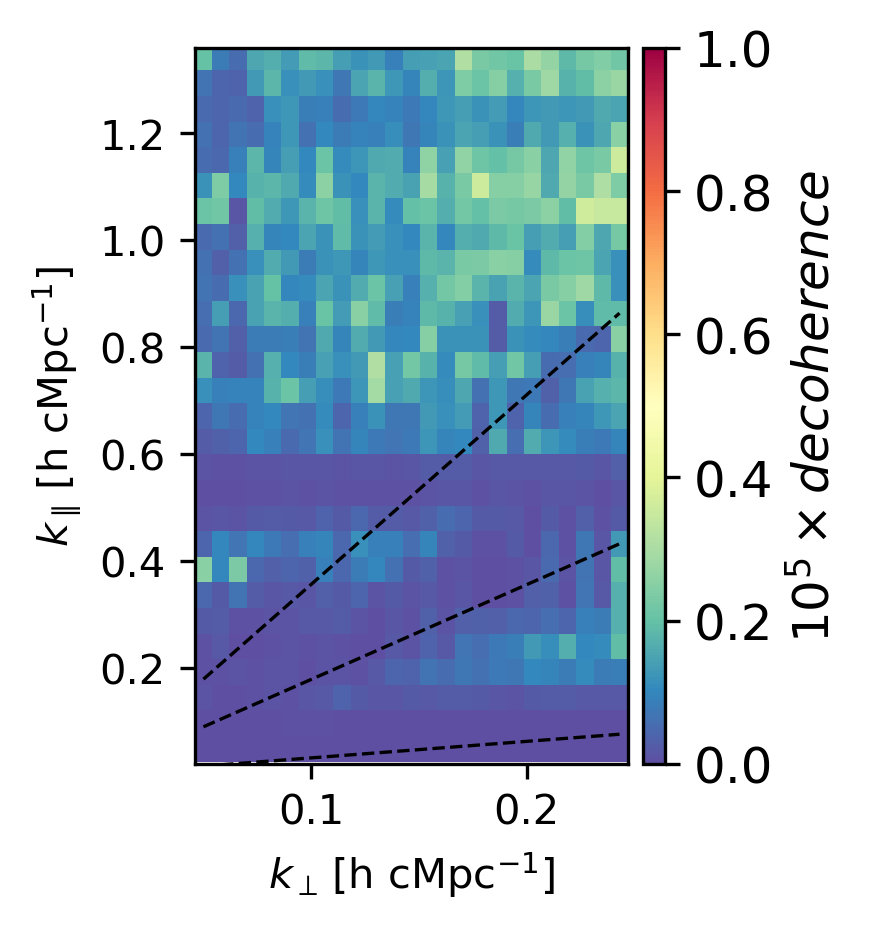}
    \caption{Amount of signal decorrelation (due to reduced $\text{S/N}$-ratio) resulting from calibration of a compressed dataset instead of its original version (without compression). The same calibration gains solutions set is applied to both the reference and compressed data before computing the coherence. The dashed lines show the 5, 30 and 90 degrees delay lines respectively.}
    \label{fig:di_compresion_noise_decoherence}
\end{figure}

\begin{figure*}
    \includegraphics[width=\linewidth]{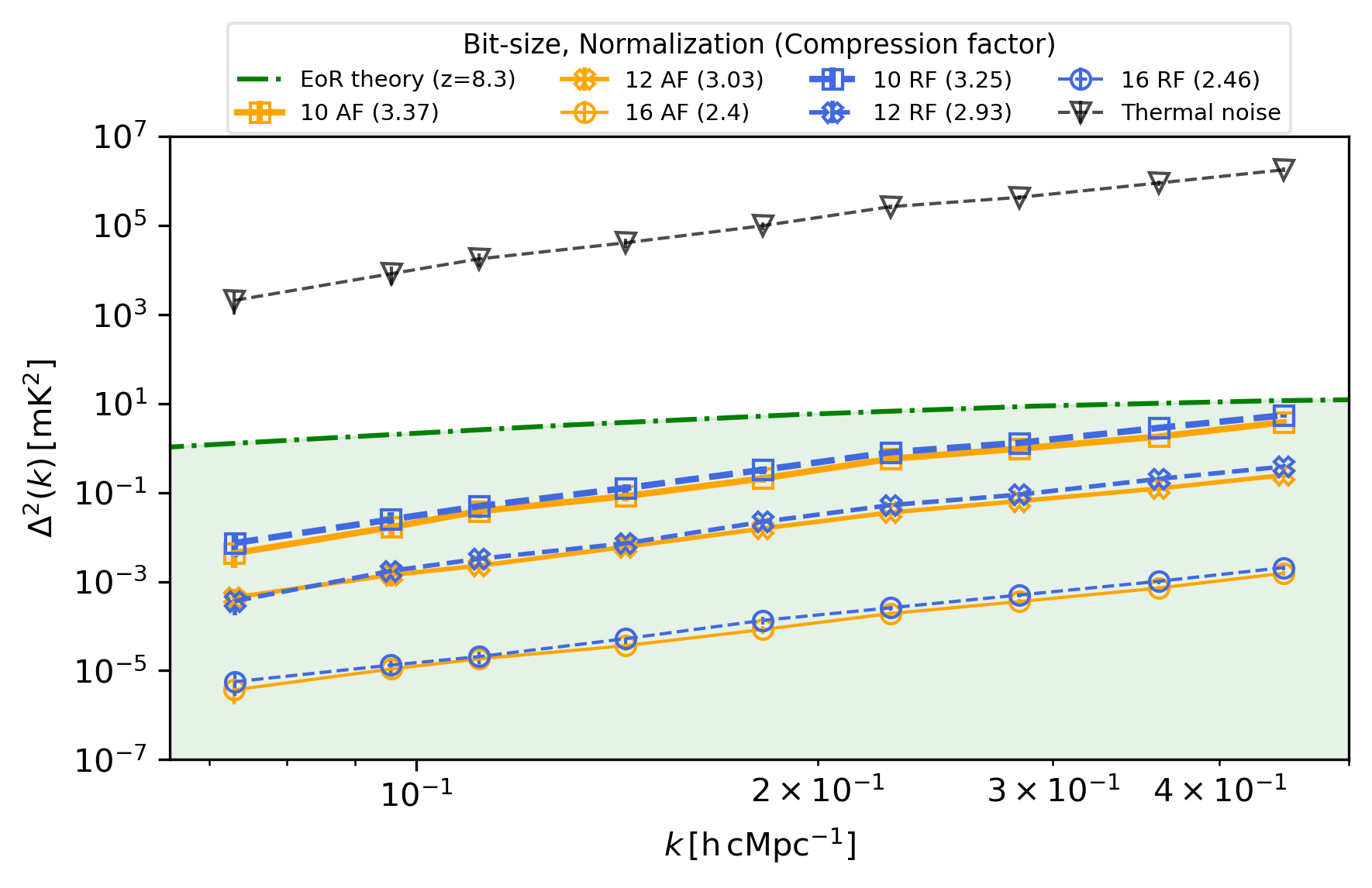}
    \caption{Comparison of the compression error levels to the spherically-averaged power spectrum and the compression factors, for different bit-sizes and normalization methods. Orange and blue lines represent AF and RF normalization methods respectively with the 3 thickness levels and markers representing the bit-sizes from 10 (thickest line with square markers) to 16 (thinnest line with circle markers). The green line shows a representative spherically-averaged power spectrum from a simulated 21-cm model at $z=8.3$. Compression errors from all settings are below the 21-cm signal even from a relatively small 6-hours dataset and since the compression noise is shown to be incoherent, a deeper data integration of say, 1000 hours, will result in result in a compression noise level that is $\times1000$ lower for all settings. The compression factors obtained for each setting are shown in brackets in the legend.
    }
    \label{fig:dysco_params_optimisation}
\end{figure*}
 
Calibration gains solutions obtained from compressed data should not show significant discrepancies from those obtained by calibrating uncompressed data. In the tests discussed above, we used gains from the reference data for the purpose of eliminating solver noise. We show that the results remained unchanged when we used the calibration gains solutions from compressed data instead. In Fig. \ref{fig:simulations_residuals_coherence}, we show the ratio of the compression residuals obtained by applying either the compressed or uncompressed data solutions. This ratio shows random fluctuations close to unity implying no significant difference.

Additionally, we examine the signal decorrelation resulting from calibrating a compressed dataset as opposed to its original uncompressed version. A unity coherence between the calibrated reference and compressed data is expected if the calibration solutions obtained for the two datasets were identical. We want to measure the level of signal decorrelation resulting from the difference in the gains. The metric is obtained from the pair of reference and compressed calibrated datasets after applying the same gains solutions set to both (either the gains obtained from the reference data or the uncompressed data\footnote{We can also apply the gains solutions obtained from calibrating the compressed dataset to the reference dataset and check the decoherence as well. The two decoherence outputs should be equivalent in the case where calibration gains solution outputs are the same regardless of whether the input data was compressed or not. The two were found to be similar to the order of $\sim10^{-7}$. Therefore, the reference solutions and \textsc{dysco} compressed solutions are almost identical. By applying either of them to both the reference and compressed data and then computing the coherence of the calibrated data pair, we obtain an equivalent estimate of the compression noise coherence.}). We use the `decoherence' metric\footnote{The actual decorrelation is due to a reduction in $\text{S/N}$-ratio  between the two calibrated datasets and not loss of the actual signal.} ($D$, equation \ref{eqn:coherence_abs_cross_ps}), a positive quantity such that an exact zero decoherence value would imply total coherence between the two calibrated datasets (i.e. no compression effect) while 1 would imply maximum decoherence:

\begin{equation}
    \label{eqn:coherence_abs_cross_ps}
    D_{a, b}^{\text{abs}} \left(k_{\perp}, k_{\parallel} \right) = 1 - \frac{ |\Tilde{T}_{a}^{*}(\mathbf{k}) \Tilde{T}_b(\mathbf{k})| }{ \sqrt{ |\Tilde{T}_a(\mathbf{k})|^2 |\Tilde{T}_b(\mathbf{k})|^2 }}.
\end{equation}

Figure \ref{fig:di_compresion_noise_decoherence} shows the decoherence resulting from this calibration of compressed data as opposed to uncompressed data due to the reduced $\text{S/N}$-ratio of the data. Here we see a decoherence at the $10^{-6}$ level. Since we have already shown that the compression noise is far below the thermal noise level, any apparent decoherence seen in the noise dominated regions (e.g at higher $k_\parallel$ modes) can be attributed to the noisy nature of the data itself in those regions as opposed to being an effect of data compression.

Nevertheless, this effect is overall insignificant and therefore calibration of compressed data yet again shows no nefarious effects.

Similarly, we found that the solver noise has a decoherence at the order of $\sim10^{-4}$, two orders of magnitude higher than the decoherence caused by data compression. This implies that the compression noise decorrelates the signal at a subdominant level in comparison to the solver noise. Compression noise is therefore not an issue of concern.

\subsection{Optimal compression settings for LOFAR-HBA data}
\label{sec:sim_results}
The error introduced by lossy compression is expected to vary depending on the compression settings chosen for Dysco. A higher bit-rate choice for the compressed data will result in a reduced compression error. Additionally, the choice of the data normalization and the quantization distribution will influence the final compression error. The previous sections used the default \textsc{dysco} settings (see Table~\ref{tab:dysco_params}) as implemented in the \textsc{DP3} pipeline. While these default parameters might be ideal for science cases such as radio surveys and transient searches, the requirements on any resulting errors in high-redshift 21-cm signal detection experiments are much more stringent. We thus intend to show how varying these compression parameters reflects on the 21-cm signal power spectrum, specifically the bit-size and normalization.
We do not test parameters that are known to be worse than the default parameters, such as the row normalization method and bit-sizes less than 10.

We also do not test the effect of different quantization distributions, firstly because any conclusions drawn from comparing different distributions would not be robust enough to apply to all datasets: and secondly, the performance of a given distribution is also coupled to other settings such as the bit-size \citep{Offringa2016}. As summarized in Table~\ref{tab:all_nights_sim} (simulation B), this test was carried out using 6-hour simulations and incorporating only the core stations of LOFAR HBA. For reference, we also included a simulated 21-cm signal model at $z=8.3$ from \cite{Mesinger2016}.

Figure \ref{fig:dysco_params_optimisation} shows the data compression factor and the spherically-averaged power spectrum of the compression residuals obtained by varying the data compression normalization method and bit-rates. Across the three tested bit-rates (10, 12, 16), both RF and AF normalization result in similar error levels that are all far below the thermal noise level. However, the RF normalization errors are about $\times 1.4$ higher than the AF errors. As described in Sect. \ref{sec:compression}, this error difference between AF and RF is attributable to the difference in the normalization dimensions between the two methods.
Nevertheless, all compression residuals are well below the theoretical EoR level at $z=8.3$ for both normalization methods, across all the bit-sizes, even in the 6 hours of data used in this test. In 1200 hours of data, for example, these errors would be 200 times lower even. 

As expected the compression noise is higher with lower bit-sizes. Moving from 10 to 12 bits results in a factor of $\sim4$ and $\sim60$ (slightly lower than the expected factor of 64) from 12 to 16, respectively. The compression factor achieved varies from 3.4, 3.0 to 2.5 for 10, 12 and 16 bit-rates, respectively. The exact compression factor is dependent on the dimensions of the data, particularly on the number of channels: the more channels stored in a measurement set, the lower the relative impact of the metadata, and the higher the compression factor. Since the tests were done on a subset of the data, the compression factor on a full LOFAR observation night dataset is $\sim4$ when done with 10 bits and AF normalization. If done with RF normalization instead, this factor will be slightly lower, however, the RF normalization method can be used to compress both visibility cross-correlations and auto-correlations while the AF normalization is limited to the cross-correlations only.

Based on these results, for LOFAR EoR data, we recommend 10-bits \textsc{dysco} compression with RF normalization. These parameters are suitable for LOFAR and might need re-tuning for other instruments. However, due to the similarities between LOFAR and SKA-Low, it suffices to conclude that the optimal parameters obtained here for LOFAR are likely also applicable for SKA-Low with minimal adjustments.

\section{Conclusions}
\label{sec:conclusions}
Lossy data compression methods can be a means to reduce the large expected costs associated with the storage and transfer of radio interferometric data, in particular those from LOFAR and the SKA. The compression should, however, not compromise the fidelity of the data, especially for high-precision studies such as 21-cm signal observations of the Epoch of Reionization and Cosmic Dawn. In this work, we have investigated the effect of lossy compression of visibilities on the 21-cm observations. Specifically, we have examined the level of compression errors and their behavior as they manifest in the 21-cm power spectrum, using the \textsc{dysco} compression code \citep{Offringa2016}.

We find that compression introduces additional noise to the power spectrum. However, this noise is around 5 orders of magnitude lower than the error on the thermal noise power spectrum of a single night. This noise also has been shown not to be correlated to the sky as seen from the minimal coherence between the residuals of different datasets.

Since the compression noise is much lower than the thermal noise and highly incoherent, its effect on calibration is found to be insignificant as expected. The calibration solutions obtained from compressed data are highly similar to those obtained from the reference data. While this test was done using only the direction-independent calibration step, it suffices to conclude that a similar insignificant effect applies in all calibration stages of the data. Thus we do not delve into compression effects in direction-dependent calibration.

We examine the optimal \textsc{dysco} compression parameters for LOFAR EoR data. We find that the bit-size used to store the compressed data is crucial in determining whether an error in the power spectrum remains well below the expected 21-cm signal after long integrations. Higher bit-sizes result in lower compression factors on one hand, but also less compression on the other. Therefore, a balance between these two factors should be considered when choosing a suitable compression bit-size. Moreover, since the compression performance depends on the instrument sensitivity, different instruments might need a retuning of these parameters. While this paper used LOFAR-HBA data, the findings reported here will likely apply to SKA-low data since both instruments will have around the same noise per visibility.

\begin{acknowledgements}
      JKC, LKV, BKG, SAB, SG, CH and SM acknowledge the financial support from the European Research Council (ERC) under the European Union’s Horizon 2020 research and innovation programme (Grant agreement No. 884760, “CoDEX”). LVEK, ARO and EC acknowledge support from the Centre for Data Science and Systems Complexity (DSSC), Faculty of Science and Engineering at the University of Groningen. F.G.M. acknowledges support from a PSL Fellowship. LOFAR, the Low-Frequency Array designed and constructed by ASTRON, has facilities in several countries, owned by various parties (each with their own funding sources), and collectively operated by the International LOFAR Telescope (ILT) foundation under a joint scientific policy. This research made use of publicly available software developed for LOFAR telescope.
\end{acknowledgements}


\bibliographystyle{aa}
\bibliography{aabib}

\begin{appendix}

\section{Coherence of the calibration solver noise}
\begin{figure}
    \includegraphics[width=\linewidth]{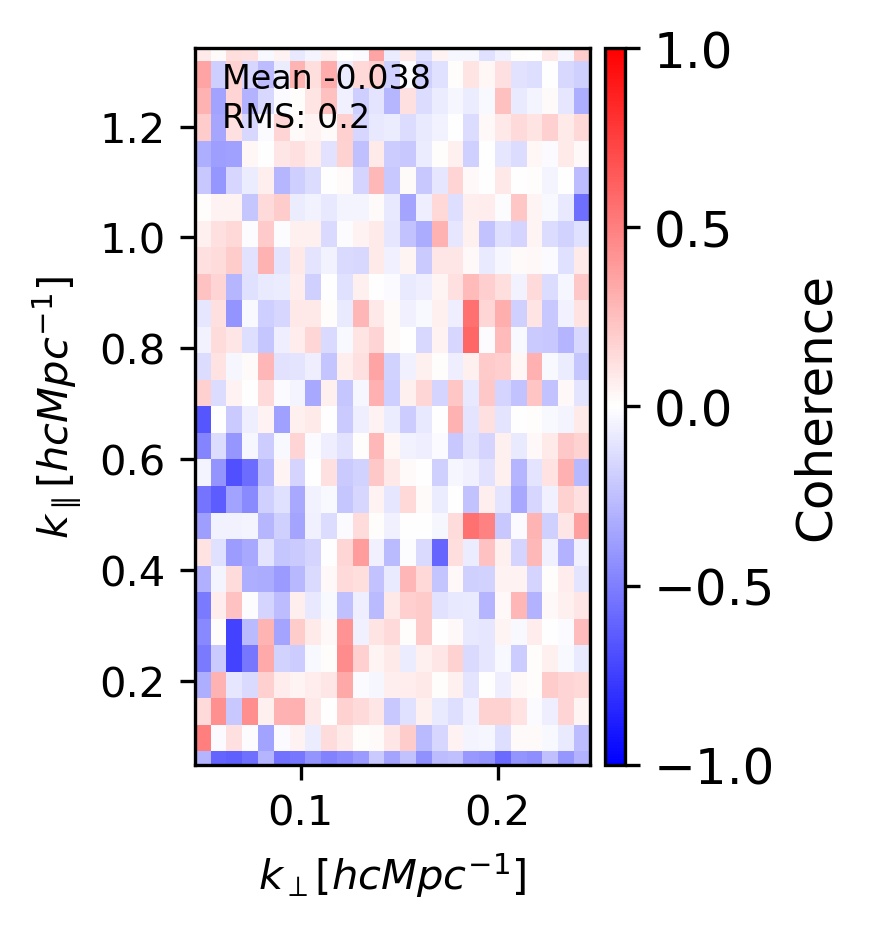}
    \caption{The ratio of compression residuals after applying either the gains solutions obtained from uncompressed data or the ones obtained from compressed data.}
    \label{fig:solver_noise_coherence}
\end{figure}

In Fig. \ref{fig:solver_noise_coherence}, we show the coherence between the solver noise obtained from the calibration runs of two different nights. Similar to the compression noise coherence shown in Fig. \ref{fig:coherence_2dps_and_average}, solver noise shows minimal coherence. Therefore, conclusions drawn from the compression noise coherence hold for solver noise as well. 
\label{app:solver_noise_coherence}

\end{appendix}
\end{document}